\documentstyle[twoside,fleqn,espcrc2]{article}

\def\al{\alpha}

\def\be{\begin{equation}}
\def\ee{\end{equation}}
\def\bea{\begin{eqnarray}}
\def\eea{\end{eqnarray}}

\def\beq{\begin{equation}}
\def\eeq{\end{equation}}
\def\bea{\begin{eqnarray}}
\def\eea{\end{eqnarray}}
\def\bq{\begin{quote}}
\def\eq{\end{quote}}

\def\al{\alpha}
\def\be{\beta}
\def\ga{\gamma}
\def\de{\delta}

\def\et{\eta}

\def\rh{\rho}

\def\si{\sigma}

\def\ph{\phi}

\def\ps{\psi}
\def\om{\omega}

\def\De{\Delta}

\def\prt{\partial}

\def\vev#1{\langle {#1}\rangle}

\def\half{{\textstyle{1\over 2}}}
\def\frac#1#2{{\textstyle{{#1}\over {#2}}}}

\def\lsim{\mathrel{\rlap{\lower4pt\hbox{\hskip1pt$\sim$}}
    \raise1pt\hbox{$<$}}}
\def\gsim{\mathrel{\rlap{\lower4pt\hbox{\hskip1pt$\sim$}}
    \raise1pt\hbox{$>$}}}
\def\sqr#1#2{{\vcenter{\vbox{\hrule height.#2pt
         \hbox{\vrule width.#2pt height#1pt \kern#1pt
         \vrule width.#2pt}
         \hrule height.#2pt}}}}

\def\rl{\stackrel{\leftrightarrow}{\hskip2pt\prt^{\mu}}}

\def\rln{\stackrel{\leftrightarrow}{\hskip2pt\prt^{\nu}}}
 
\def\AJ{{Ap. J.} }

\def\APP{{Acta Phys. Pol.} }

\def\APP{{Astropart. Phys.} }

\def\CQG{{Class. Quantum Gravity} }

\def\JP{{J. Phys.} }

\def\NAT{{Nature} }

\def\NP{{Nucl. Phys.} }
\def\PL{{Phys. Lett.} }
\def\PR{{Phys. Rev.} }
\def\PRL{{Phys. Rev. Lett.} }
\def\PRTS{{Phys. Rep.} }

\def\PMAG{{Philos. Mag.} }

\def\gappeq{\mathrel{\rlap {\raise.5ex\hbox{$>$}}
{\lower.5ex\hbox{$\sim$}}}}
 
\def\lappeq{\mathrel{\rlap{\raise.5ex\hbox{$<$}}
{\lower.5ex\hbox{$\sim$}}}}

\title{On the spontaneous breaking of Lorentz invariance}

\author{Orfeu Bertolami
\address{Departamento de F\'{\i}sica, Instituto Superior T\'ecnico, \\
Av. Rovisco Pais, 1049 - 001 Lisboa, Portugal}}

\begin{document}

\begin{abstract}

We show that estimates of the Lorentz symmetry violation 
extracted from ultra-high energy cosmic rays beyond the GZK cut-off 
set bounds on the parameters of a Lorentz-violating extension 
of the Standard Model. Moreover, we 
argue that correlated measurements of the difference in the
arrival time of gamma-ray photons and neutrinos emitted from
Active Galactic Nuclei or Gamma-Ray Bursts may provide a signature
for a possible violation of the Lorentz symmetry. We find
that this time delay is energy independent, but that it has a dependence 
on the chirality of the particles involved. 

\end{abstract}
\maketitle
\section{INTRODUCTION}

Lorentz invariance is one of the most fundamental symmetries of physics and 
is an underlying ingredient of all known physical theories. However, more 
recently, there has been theoretical evidence that, in the realm of 
string/M-theory, this symmetry may be violated. 
This naturally poses the question of verifying this violation 
experimentally. 
In this contribution\footnote{Invited 
talk delivered at the Third Meeting on Constrained Dynamics and 
Quantum Gravity, Villasimius, Sardinia, September 1999.}, 
we study the implications of a putative 
violation of Lorentz invariance in the context of a Lorentz-violating 
extension of the Standard Model (SM) aiming to analyse its impact 
on the physics of ultra-high energy cosmic rays and the possibility of 
an astrophysical test of this violation \cite{Bertolami}.  

Actually, in what concerns the latter issue, it is striking that 
there is convincing evidence 
that the observed jets of 
Active Galactic Nuclei (AGN) are efficient cosmic proton accelerators.   
Furthermore, the photoproduction of neutral pions by accelerated protons 
is assumed to be
the source of the highest-energy photons through which most of the luminosity
of the galaxy is emitted. The decay of charged pions with the ensued
production of neutrinos is another distinct signature of the proton induced
cascades and estimates of the neutrino flux are 
believed to be fairly model independent \cite{Mannheim,Waxman1}. 
Gamma-Ray Bursts (GRB) have been also 
suggested as a possible source of high-energy neutrinos \cite{Waxman2}. 
Moreover, a deeper understanding of these sources is expected as 
large area ($\sim km^2$) high-energy neutrino telescopes are under 
construction (see e.g. \cite{Gaisser}). These telescopes will allow 
obtaining information 
that is congenially correlated with gamma-ray flares and bursts
emitted by AGN and GRB sources. On the other hand, it has already 
been pointed out that astrophysical observations
of faraway sources of gamma radiation could provide important hints
on the nature of gravity-induced effects 
\cite{Amelino1,Biller,Ellis} and hence on physics beyond the 
Standard Model (SM). 
We argue that delay measurements 
in the arrival time of correlated sources of gamma radiation 
and high-energy neutrinos can, when considered in the context of 
a Lorentz-violating extension of the SM \cite{Colladay}, 
help setting relevant
limits on the violation of that fundamental symmetry.  
As we shall see, we can relate our results with the recently discussed 
limit on the violation of Lorentz symmetry from the observations of
high-energy cosmic rays beyond the GKZ cut-off \cite{Coleman}.

The idea of dropping the Lorentz symmetry has been 
sugested long ago. Indeed, a background 
cosmological vector field has been considered as a  
way to introduce our velocity with respect to a preferred frame of reference 
into the physical description \cite{Phillips}. It has also been
proposed, based on the behaviour of the renormalization group
$\beta-$function of non-abelian gauge theories,
that Lorentz invariance could be actually a low-energy symmetry
\cite{Nielsen}. In higher dimensional theories of gravity, 
models that are not locally Lorentz invariant have been  
studied in order to obtain light fermions in chiral representations
\cite{Weinberg}.

The spontaneous breaking of Lorentz symmetry due to
non-trivial solutions of string field theory was first discussed
in Refs. \cite{Kostelecky1}. These non-trivial solutions
arise in the context of the string field theory of open strings and may
have striking implications at low-energy. For example, assuming that 
the contribution 
of Lorentz-violating interactions to the vacuum energy is about half of 
the critical density leads to the conclusion that 
quite feeble tensor mediated interactions
in the range of about $10^{-4}~m$ should exist \cite{Bertolami1}. 
Lorentz violation may lie, with the help of inflation, at the origin
of the primordial magnetic fields 
required to explain the observed galactic magnetic field as it 
may also imply in the breaking of conformal symmetry of electromagnetism 
\cite{Bertolami2}.
It is quite natural that violations of the Lorentz invariance may imply 
in the breaking of CPT symmetry \cite{Kostelecky}. 
Interestingly, this  
possibility can be verified experimentally in 
neutral-meson \cite{Kostelecky2} 
experiments, Penning-trap 
measurements \cite{Bluhm1} and hydrogen-antihydrogen spectroscopy 
\cite{Bluhm2}. 
Moreover, the breaking of CPT symmetry also allows for 
an explanation of the baryon asymmetry of the Universe, as 
tensor-fermion-fermion interactions expected in the low-energy limit of 
string field theories give rise 
to a chemical potential that creates in equilibrium a
baryon-antibaryon asymmetry in the presence of baryon number 
violating interactions \cite{Bertolami3}.

Limits on the violation of Lorentz symmetry have been 
directly investigated through
laser interferometric versions of the Michelson-Morley experiment  
where comparison between the velocity of light, $c$, and the  
maximum attainable velocity of massive particles, $c_i$, up to  
$\delta \equiv |c^2/c_{i}^2 - 1| < 10^{-9}$ \cite{Brillet}. In 
the so-called Hughes-Drever experiment \cite{Hughes,Drever} much more 
stringent limits can be obtained, searching for a time dependence 
of the quadrupole splitting of nuclear Zeeman levels
along Earth's orbit, e.g.
$\delta < 3 \times 10^{-22}$ \cite{Lamoreaux}. Impressively, a more 
recent assessment of these experiments reveals that more accurate 
bounds, up to 8 orders 
of magnitude, can be reached \cite{Kostelecky3}.
From the astrophysical side, limits on the violation of momentum 
conservation and the 
existence of a preferred reference frame can also be established from 
bounds on the
parametrized post-Newtonian parameter, $\alpha_{3}$. This parameter vanishes 
in General Relativity and can be extracted 
from the pulse period of pulsars and millisecond  
pulsars \cite{Bell}. The most 
recent bound, $|\alpha_{3}| < 2.2 \times 10^{-20}$ \cite{BellD}, 
indicates that Lorentz symmetry is unbroken up to that level.

In the next sections, we shall calculate the corrections to 
the dispersion relation
arising from a Lorentz-violating extension of the SM and 
confront it with the evidence on the violation of Lorentz 
invariance arising from the ultra-high energy cosmic ray physics. We shall 
also see 
that these corrections induce a time delay in the arrival of signals from 
faraway sources carried by different particles.

\section{LORENTZ-VIOLATING EXTENSION OF THE STANDARD MODEL}

It is widely believed that SM is a 
low-energy description of a more fundamental theory, where all 
interactions including gravity are unified and the 
hierarchy problem is solved. It is quite plausible 
that, in this most likely higher-dimensional fundamental theory, 
symmetries such as 
CPT and Lorentz invariance, may undergo spontaneous symmetry breaking. 
The fact that within string/M-theory, currently the most promising candidate 
for a fundamental theory, a mechanism where spontaneous breaking of 
Lorentz symmetry is known \cite{Kostelecky1,Kostelecky,Dvali}, 
suggests that the violation of those symmetries might actually occur and that 
its implications should be investigated.

There is no reason, at least in principle, for this breaking not to 
extend into the
four-dimensional spacetime. If this is indeed the case, CPT and Lorentz
symmetry violations will be likely to take place within the SM and its 
effects might be detected.
In order to account for the CPT and Lorentz-violating effects an extension 
to the minimal SM has been developed 
\cite{Colladay} based on the assumption that CPT and Lorentz-violating terms
might arise from interactions of tensor fields to Dirac fields 
when Lorentz tensors acquire non-vanishing vacuum expectation 
values. Interactions of this form are expected to arise from the 
string field trilinear self-interaction, as is in the 
open string field theory \cite{Kostelecky1,Kostelecky}. 
These interactions may also emerge from the scenario where our world 
is wrapped in a $3$-brane and this is allowed to tilt \cite{Dvali}.
In order to preserve the SM power-counting renormalizability 
only terms involving operators of mass dimension four
or less are considered in this extention. In \cite{Bertolami}, 
only the fermionic sector of the extension \cite{Colladay} 
was considered~\footnote{It was assumed that SM gauge sector is unaltered. 
Changing this sector has been already 
considered, but the phenomenological restrictions are  
quite severe, 
at least in what concerns the term that gives origin to a cosmological 
birefringence \cite{Carroll}.}. 
This sector includes both leptons and quarks, since
SU(3) symmetry ensures violating extensions to be colour-independent. The
extended fermionic sector consists of CPT-odd and CPT-even contributions, 
which are given by \cite{Colladay}
\beq
{\cal L}^{\rm CPT-odd}_{\rm Fermion} = - a_{\mu} \overline{\psi} 
\gamma^{\mu} \psi -
b_{\mu} \overline{\psi} \gamma_5 \gamma^{\mu} \psi \quad,
\label{1.1}
\eeq
\begin{eqnarray}
{\cal L}^{\rm CPT-even}_{\rm Fermion} &=& \half i c_{\mu\nu} \overline{\psi}
\gamma^{\mu} \rl \psi \nonumber\\
&+& \half i d_{\mu\nu} \overline{\psi} \gamma_5
\gamma^{\mu} \rl  \psi \nonumber\\
&-& H_{\mu\nu} \overline{\psi}
\sigma^{\mu\nu} \psi~~,
\quad
\label{2.1}
\end{eqnarray}
where the coupling coefficients $a_{\mu}$ end $b_{\mu}$ have dimensions of
mass, $c_{\mu\nu}$ and $d_{\mu\nu}$ are dimensionless and can have both
symmetric and anti-symmetric components, and $H_{\mu\nu}$ 
has dimension of mass and is anti-symmetric. All Lorentz violating 
parameters are hermitian and flavour-dependent. Some of them 
may induce flavour changing 
neutral currents when non-diagonal in flavour.
  
The Langrangian density of the fermionic sector including
Lorentz-violating terms reads:
\begin{eqnarray}
{\cal L} &=& \half i \overline{\psi}~\gamma_{\mu} \rl \psi - 
a_{\mu} \overline{\psi}~\gamma^{\mu} \psi - b_{\mu} \overline{\psi} 
~\gamma_5 \gamma^{\mu} \psi \nonumber\\
&+& \half i c_{\mu\nu} \overline{\psi}~\gamma^{\mu} \rln\psi 
+ \half i d_{\mu\nu} \overline{\psi}~\gamma_5 \gamma^{\mu} \rln \psi 
\nonumber\\
&-& H_{\mu\nu} \overline{\psi}~\sigma^{\mu\nu} \psi - m \overline{\psi} 
\psi ~~,
\label{2.2}
\end{eqnarray}
where only kinetic terms are kept as we are interested in
deducing the free particle energy-momentum relation. 

From the above Lagrangian density, we can get the Dirac-type
equation
\begin{eqnarray}
&&[i \gamma^{\mu} (\prt_{\mu} + (c_{\mu}^{~\al} - d_{\mu}^{~\al} 
\gamma_5)~
\prt_{\al}) - a_{\mu} \gamma^{\mu} \nonumber\\
&&- b_{\mu} \gamma_5 \gamma^{\mu} 
- H_{\mu\nu} \sigma^{\mu\nu} -m] \psi = 0 \quad.
\label{2.3}
\end{eqnarray}

In order to obtain the corresponding Klein-Gordon equation, we multiply eq.
(\ref{2.1}) from the left by itself with
an opposite mass sign yielding:
\begin{eqnarray}
&&\biggl[[i (\prt_{\mu} + c_{\mu}^{\al} \partial_{\al}) - a_{\mu}]^2 
+(d_{\mu}^{\al} \prt_{\al})^2 - b^2
- m^2 \nonumber \\
&-& i \si^{\mu\rh} [i \prt_{\mu} c_{\rho}^{~\beta} \prt_{\beta}
+ i c_{\mu}^{~\alpha} \prt_{\alpha} i \prt_{\rho} 
+ i c_{\mu}^{~\alpha} \prt_{\alpha} i c_{\rho}^{~\beta} \prt_{\beta}
\nonumber \\  
&-& i (\prt_{\mu} + c_{\mu}^{~\alpha} \prt_{\alpha}) i d_{\rho}^{~\beta} 
\ga_5 \prt_{\beta} + i d_{\mu}^{~\alpha} \ga_5 \prt_{\alpha}
i (\prt_{\rho} \nonumber \\
&+& c_{\rho}^{~\beta} \prt_{\beta})
- 2 b_{\mu} \ga_5 [i (\prt_{\rh} + c_{\rh}^{\be} \prt_{\be}) - a_{\rh}]]
\nonumber \\
&-& 2 i (i a_{\mu} \si^{\mu\rh} - b_{\mu} \ga_5 g^{\mu\rh}) 
d_{\rh}^{~\be} \ga_5 \prt_{\be} \nonumber \\  
&+& \si^{\mu\nu} \si^{\rh\si} H_{\mu\nu} H_{\rh\si} 
- H_{\rh\si}(\ga^{\mu} \si^{\rh\si}
+ \si^{\rh\si} \ga^{\mu})  \nonumber \\
&&[i (\prt_{\mu} 
+ (c_{\mu}^{~\al} - d_{\mu}^{~\al} \ga_5) \prt_{\al})  \nonumber \\
&-& a_{\mu} + b_{\mu}
\ga_5]\biggr] \ps = 0 \quad.
\label{2.4}
\end{eqnarray}
To eliminate the off-diagonal terms, the squaring procedure has to
be repeated once again. However, since Lorentz symmetry breaking 
effects are quite constrained experimentally,  
violating terms higher than second order will be dropped. After some algebra, 
we find that off-diagonal terms cannot be fully cancelled, but that 
these terms are higher order in the Lorentz violating parameters. To 
simplify further
we also drop $H_{\mu\nu}$. This is justifiable as 
only time-like Lorentz-violating parameters are going to be studied. Hence,
we obtain, for the Klein-Gordon type equation, up to second 
order in the new parameters:
\begin{eqnarray}
&&\biggl[[(i \prt)^2 + 2i \prt_{\mu} i c^{\mu\al} \prt_{\al} - 2i \prt_{\mu}
a^{\mu} - m^2]^2 \nonumber \\
&+&4 i \prt_{\mu} i d_{\rho}^{~\be} \prt_{\beta} i \prt_{\eta}
i d_{\phi}^{~\delta} \prt_{\delta} (g^{\mu \et} 
g^{\rh\ph} - g^{\mu\rh} g^{\et\ph})
\nonumber \\  
&-& 8 i \prt_{\mu} i d_{\rho}^{~\be} \prt_{\beta} b_{\eta} i \prt_{\phi}
(g^{\mu \rho} g^{\eta\ph} - g^{\mu\phi} g^{\rho\eta})\nonumber \\
&+& 4 b_{\mu} b_{\et} i \prt_{\rh} i \prt_{\ph} (g^{\mu \et}
g^{\rh\ph} - g^{\mu\rh} g^{\et\ph}) \biggr] \ps = 0~~.
\label{2.5}
\end{eqnarray}
Thus, in momentum space we get at 
lowest non-trivial order, the following relationship:
\begin{eqnarray}
&&(p_{\mu} p^{\mu} + 2 p_{\mu} c^{\mu\al} p_{\al} + 2 p_{\mu} a^{\mu} -
m^2)^2 \nonumber \\
&+& 4 [p_{\mu} p^{\mu}  d_{\rho}^{~\be} p_{\beta} d^{\rho \delta} p_{\delta} 
- (p_{\mu} d^{\mu \beta} p_{\beta})^2]
\nonumber \\
&+& 8 (p_{\mu} p^{\mu}  d_{\eta}^{~\be} p_{\beta} b^{\eta} -
p_{\mu} d^{\mu\be} p_{\beta} b_{\eta} p^{\eta})\nonumber \\
&+& 4 [b_{\mu} b^{\mu} p_{\nu} p^{\nu} - (b_{\mu} p^{\mu})^2] = 0~~.
\label{2.6}
\end{eqnarray}
Hence, the dispersion relation arising from the 
Lorentz-violating extension of the SM is the following:
\begin{eqnarray}
&&p_{\mu} p^{\mu} - m^2 = -2 p_{\mu} c^{\mu\al} p_{\al} - 2 p_{\mu} a^{\mu}  
\nonumber \\
&\pm& 2 \biggl[(p_{\mu} d^{\mu \beta} p_{\beta})^2 - p_{\mu} p^{\mu} 
d_{\eta}^{~\be} p_{\beta} d^{\eta \delta} p_{\delta}
\nonumber \\
&+&2 (p_{\mu}  d^{\mu\be} p_{\beta} b_{\rho} p^{\rho} - 
p_{\mu} p^{\mu} d_{\eta}^{~\be} p_{\beta} b^{\eta})\nonumber \\
&-& b_{\mu} b^{\mu} p_{\nu} p^{\nu} + (b_{\mu} p^{\mu})^2
\biggr]^{1/2} \quad,
\label{2.7}
\end{eqnarray}
where the $\pm$ sign refers to the fact that the effects of $b_{\mu}$ and 
$d_{\mu\nu}$ depend on chirality. 

Finally, we consider, for simplicity, the scenario where coefficients 
$a_{\mu}$, $b_{\mu}$, $c_{\mu\nu}$ and $d_{\mu\nu}$ have only time-like 
components, which yields the simplified dispersion relation 
\beq
p_{\mu} p^{\mu} - m^2 = - 2 c_{00} E^2 - 2 a E \pm 2 (b + d_{00}E) p~,
\label{2.8}
\eeq
where we have dropped the component indices of coefficients $a$ and $b$.
From now on we drop parameter $a$ as it
is potentially dangerous from the point of view of giving origin to 
flavour changing neutral currents when more than one flavour is involved. 

In the next section, we shall use the dispersion relation 
(\ref{2.8}) to examine how
GZK cut-off for ultra-high energy cosmic rays can be relaxed. The 
ensued discussion is similar to the one described in \cite{Coleman}, 
where it is assumed that the limiting velocities of particles 
in different reference frames are {\it ad hoc} different.

\section{ULTRA-HIGH ENERGY COSMIC\\ RAYS}

The discovery of the cosmic background radiation made inevitable
the question of how the most energetic cosmic-ray particles would be
affected by the interaction with the microwave photons. Actually,
the propagation of the ultra-high energy nucleons is limited
by inelastic impacts with photons of the background
radiation so that nucleons with energies above $5 \times 10^{19}~eV$
are unable to reach Earth from further than $50 - 100~Mpc$. 
This is the well known
GZK cut-off \cite{Greisen}. However, events where the estimated energy of 
the cosmic primaries is beyond the GZK cut-off have been observed 
by different collaborations \cite{Yoshida,Bird,Brooke,Efimov}. 
It has been suggested \cite{Coleman} (see also \cite{Mestres}) 
that slight violations of Lorentz
invariance could be at the origin of energy-dependent effects 
which would suppress processes,
otherwise dynamically inevitable, such as for instance 
the resonant reaction,
\beq
p + \ga_{2.73K} \to \De_{1232} \quad,
\label{3.1}
\eeq
which is central to the GZK cut-off. 
Were this process untenable, the
GZK cut-off would not hold and therefore a cosmological origin for the
high-energy cosmic radiation is theoretically acceptable. 
As discussed in \cite{Coleman},
this can occur through a change in the dispersion relation for free particles.
We shall see that this is indeed what happens when process
(\ref{3.1}) is analysed with dispersion relation (\ref{2.8}).  
Considering a head-on impact of a proton of energy $E$ with a 
cosmic background radiation photon of energy $\om$, the likelihood of the 
process (\ref{3.1}) would be conditioned on satisfying 
(cf. eq. (\ref{2.8}))  
\beq
\label{eq}
2\om + E \ge  m_{\Delta} (1 -  c^{\Delta}_{00})\quad.
\label{3.2}
\eeq
Thus, we get from (\ref{2.8}) 
after squaring (\ref{3.2}) and dropping the $\omega^2$ term
\beq
2\om + {m_{p}^2 \over 2 E} \ge (c^{p}_{00} - c^{\Delta}_{00}) E  
+ {m_{\De}^2 \over 2 E} \quad,
\label{3.3}
\eeq
which clearly exhibits Lorentz-violating terms.

Let us now compare (\ref{3.3}) with the results of Ref. \cite{Coleman} and
show that this leads to a bound on $\Delta c_{00}$. In order to 
modify the usual 
dispersion relation for free particles, Coleman and Glashow suggested 
assigning a maximal attainable velocity to each particle. 
Thus, for a given particle $i$ moving freely in the preferred frame, 
which could be thought of as the
one in relation to which the cosmic background radiation is isotropic,
the dispersion relation would be
\beq
E^2 = p^2 c_{i}^2 + m_{i}^2 c_{i}^4 \quad.
\label{3.4}
\eeq 
Hence, the likelihood of the process (\ref{3.1}) to occur
under the conditions described above would depend on 
satisfying the kinematical requirement $2\om + E \ge m_{eff}$, 
where the effective mass $m_{eff}$ is given by
\beq
m_{eff}^2 \equiv m_{\Delta}^2 - 
(c_{p}^2 - c_{\De}^2) p^2 \quad,
\label{3.5}
\eeq
the momentum being with respect to the preferred frame.

The likelihood condition takes then the following form
\beq
2\om + {m_{p}^2 \over 2 E} \ge (c_{p} - c_{\De})E 
+ {m_{\De}^2 \over 2 E} \quad,
\label{3.7}
\eeq
where the term proportional to $c_{p} - c_{\Delta}$ is clearly 
Lorentz-violating. If the difference in the maximal velocities exceeds 
the critical value 
\beq
\de (\om) = {2 \om^2 \over m_{\De}^2 - m_{p}^2} \quad,
\label{3.8}
\eeq
then reaction (\ref{3.1}) would be forbidden and consequently the GZK cut-off
relaxed. For photons of the microwave background, $T = 2.73~K$, and 
$\omega_{0} \equiv kT = 2.35 \times 10^{-4}~eV$, this condition would be
\beq
c_{p} - c_{\De} = \de (\om_{0}) \simeq 1.7 \times 10^{-25} \quad,
\label{3.9}
\eeq
which is a quite impressive bound on the violation of the Lorentz symmetry, 
even though valid only for the process in question. Similar bounds for 
other particle pairs, although less stringent, were discussed in 
\cite{Coleman,Halprin}. 

Finally, comparison of (\ref{3.7}) with (\ref{3.3}) yields: 
\beq
c^{p}_{00} - c^{\Delta}_{00} \simeq 1.7 \times 10^{-25}  \quad.
\label{3.10}
\eeq
Thus, we see that the Lorentz-violating extension of the SM can also describe
the phenomenology of ultra-high energy cosmic rays and explain the 
violation of the GZK cut-off. Of course, the situation would be more 
complex if the Lorentz-violating parameters were allowed to have space-like 
components which would lead to direction and helicity-dependent effects.

\section{AN ASTROPHYSICAL TEST OF LO\-RENTZ INVARIANCE}

We now turn to the discussion of a possible astrophysical test of Lorentz
invariance. From eq. (\ref{3.10}) we see that
$\Delta c_{00} \simeq \epsilon$, where $\epsilon$ is a small constant 
specific of the process involved (cf. eqs. (\ref{3.9}) and (\ref{3.10})) 
and, for instance, $\vert \epsilon \vert \lsim few \times 10^{-22}$ 
from the search of neutrinos oscillations \cite{Brucker,Glashow}. 
As far as signals simultaneously emitted by faraway sources are concerned, 
the resulting effect in the propagation velocity of 
particles with energy, $E$, and momentum, $p$, is given 
in the limit where $m << p, E$ or, for massless 
particles, by $c_i = c [1 - (c_{00} \pm d_{00})_{i}]$. 
Therefore, for sources at a distance $D$, the delay in the arrival 
time will be given by:
\beq
\Delta t \simeq {D \over c} [(c_{00} \pm d_{00})_{i} - 
(c_{00} \pm d_{00})_{j}] \equiv \epsilon_{ij}^{\pm} {D \over c}~~,
\label{4.1}
\eeq
where a new constant, $\epsilon_{i j}^{\pm}$, 
involving a pair of particles is defined. 
This time delay may, despite being given by the 
difference between two fairly small numbers, be measurable for sufficiently 
far away sources. Furthermore, our result shows that the estimated 
time delay is energy independent, in 
opposition to what was obtained from general arguments 
\cite{Amelino1,Ellis}. We have also found that the time delay has a 
dependence on the chirality of the particles involved. 
In the next section we shall discuss how to estimate 
the observational value of $c^{i}_{00}$ 
(and $d^{i}_{00}$ if $d^{i}_{00} \sim c^{i}_{00}$).   

Therefore if, for instance, the signals from faraway sources were, 
as suggested previously in the introduction, from AGN $TeV$ gamma-ray flares 
and the intrinsically
related neutrino emission, we should expect for the time delay, 
\beq
\Delta t \simeq (c_{00} \pm d_{00})_{\nu} {D \over c} \quad,
\label{4.2}
\eeq
if as argued above the photon propagation is unaltered and 
assuming that the neutrinos are massless, an issue that will be most
probably 
settled experimentally in the near future. It is worth stressing 
that even before that, the effect of neutrino masses and other 
intrinsic effects related with the nature of the neutrino 
emission can, 
at least in principle, be extracted from 
data of several correlated detections of $TeV$ gamma-ray 
flares and neutrinos, if a systematic delay of 
neutrinos is observed. 
Moreover, the available knowledge of AGN phenomena and the 
confidence on the astrophysical methods available to determine their 
distance from us, make it plausible 
that the time delay strategy may
provide relevant bounds on the
violation of Lorentz symmetry. Of course, similar arguments may 
equally apply to GRB, however, the lack of
a deeper understanding of these transient phenomena introduces further 
undesirable uncertainty. It is also important to
point out that bounds involving photon and neutrinos are currently unknown and
that a difference in the arrival time between neutrinos and antineutrinos 
is expected if $d^{i}_{00}$ is non-vanishing.

\section{CONCLUSIONS}
 
In summary, we have shown that parameters of the Lorentz-violating
extension of the SM proposed in Ref. \cite {Colladay} can be related with the
phenomenology of ultra-high energy cosmic rays with the conclusion that,
as in \cite{Coleman}, it may lead to the suppression of processes responsible
for the GZK cut-off. This is a crucial argument in favour of a
possible extra-galactic 
origin for the ultra-high energy cosmic rays. We have also found that 
the relevant Lorentz-violating parameter is, at high energies, $c^{i}_{00}$ 
so that $\Delta c_{00} \simeq \epsilon$ with 
$\vert \epsilon \vert \lsim few \times 10^{-22}$ from neutrino physics 
and $\vert \epsilon \vert \lsim 10^{-25}$ from the ultra-high energy
cosmic ray physics.
It is possible to estimate the typical scales assuming that the 
source of Lorentz symmetry violation is due to non-trivial solutions 
in string field theory. Indeed, these
solutions imply that Lorentz tensors acquire vacuum expection values as  
Lorentz symmetry is spontaneously broken due to string induced 
interactions \cite{Kostelecky1,Kostelecky}. 
A parametrization for these expectation values and, hence, for $c^{i}_{00}$ 
would be for a fixed energy scale, E, the following 
\cite{Kostelecky,Bertolami3}:
\beq
c^{i}_{00} \simeq {\vev{T} \over M_{S}} = 
\lambda_{i} \left({m_{L} \over M_{S}}\right)^{l} 
\left({E \over M_{S}}\right)^{k}  \quad,
\label{6.1}
\eeq
where $T$ denotes a generic Lorentz tensor, $\lambda_{i}$ is presumably
an order one flavour-dependent constant\footnote{A scenario were $\lambda_{i}$
is of order of the respective Yukawa coupling has been also discussed 
\cite{Kostelecky}.}, $m_{L}$ is a light 
mass scale, $M_{S}$ is a string scale presumably close to Planck's mass 
or a few orders 
of magnitude below it, and $k, l$ are 
integers labelling the order of the string corrections at low-energy. 
Thus, in the
lowest non-trivial order, $k = 0,~l = 1$ ($k=l=0$ being already 
excluded experimentally), $c^{i}_{00} = \lambda_{i}
\left({m_{L} \over M_{S}}\right)$ and 
different $\lambda_{i}$ constants lead to 
$\epsilon \simeq \left({m_{L} \over M_{S}}\right)$.
Thus, we see that theoretical estimates are consistent
with high-energy cosmic ray data. Moreover, if for example, 
$\epsilon \lsim 10^{-23}$,
then it follows that $m_{L} \sim 10^{2}~KeV$ for $M_{S} \simeq M_P$ or 
$m_{L} \sim 10^{2}~eV$ if $M_{S} \simeq few \times 10^{16}~GeV$ \cite{Witten}.
Estimates for $m_{L}$ would clearly change by many orders of magnitude if 
$\lambda_{i}$ were of order of the Yukawa coupling. In either case, we 
can conclude that choice  $k =0,~l = 1$ implies the time delay 
in the arrival 
of signals from faraway sources is energy independent.
We have found, however, an interesting dependence on the chirality of 
particles involved. 

Of course, another scenario would emerge from a different choice of integers
$k, l$. For instance, the choice $k=2$ and $l=0$, the relevant one in the 
CPT symmetry violating baryogenesis
scenario \cite{Bertolami3}, where the energy should, in this case, 
be related with 
the early Universe temperature. This would imply that the time delay in the 
arrival of signals from faraway sources 
would be proportional to the square of the energy. This choice would also 
lead to the
conclusion that Lorentz violating effects, whether due to string physics 
or quantum gravity, are quadratic in the energy. Similar conclusions 
concerning the order of quantum gravity low-energy 
effects are obtained from the study of corrections to the 
Schr\"odinger equation arising from quantum cosmology in the minisuperspace
approximation \cite{Bertolami4}.    
   
In another theoretical setting, the spontaneous 
breaking of the Lorentz symmetry
may occur is the so-called braneworld \cite{Dvali}. In this scenario, 
SM particles lie on a 
$3$-brane, $\phi(x)$, embedded in spacetime, with possibly 
large compact extra dimensions, 
whereas gravity propagates in the bulk. Thus, a tilted brane would 
induce rotational 
and Lorentz non-invariant terms in the four-dimensional effective theory
as brane-Goldstones couple to all particles on the brane via an induced metric
on the brane. This will lead to operators of the form:
\beq
\prt_{\mu} \phi~\prt_{\nu} \phi~\overline{\psi}~ 
\gamma^{\mu}  \prt^{\nu} \psi + ... \quad,
\label{6.2}
\eeq  
which clearly resemble some of the Lorentz-violating terms in the SM extention 
discussed above. 
Corrections to the kinetic terms of gauge fields are also expected.  
As before, phenomenology sets tight constraints on this scenario remaining, 
however, unable to establish whether the 
breaking of Lorentz invariance, if
observed at all, has its origin on the non-perturbative nature of branes 
or has its roots in 
the perturbative string field theory scenario described above.  
The former possibility should more probably be associated with a 
$M_S$ scale that 
is a few orders of magnitude below Planck scale, while the latter to a value of
$M_S$ that should be associated with the Planck scale itself.

Finally, we have outlined a strategy 
to establish to what extent Lorentz invariance is violated,
from the observation of the 
time delay in the detection of $TeV$ gamma-ray flares and neutrinos from
AGN. Our analysis reveals that the time delay has a dependence on the 
chirality of the particles involved, but is energy independent, 
contrary to what one could expect from general arguments. In either case, 
if ever observed, a time delay in the arrival of signals from 
faraway sources would be a strong 
evidence of new physics beyond the SM.

I would like to thank Carla Carvalho for collaboration in the project that
gave origin to this contribution and Alan Kosteleck\'y for the relevant 
comments and suggestions. I am also extremely grateful to the  
organizing committee of the Constrained Dynamics and Quantum 
Gravity 1999 for the superb organizational work and for the 
most warm hospitality with which myself and my family were received in 
Sardegna.  



\begin{thebibliography}{9}
 
\bibitem{Bertolami} O.\ Bertolami and C.S.\ Carvalho, Proposed
astrophysical 
test of Lorentz invariance, gr-qc/9912117; to appear in Phys. Rev. D.
  
\bibitem{Mannheim} K.\ Mannheim, \APP 3 (1995) 295;
   
F.\ Halzen and E.\ Zas, \AJ 488 (1997) 669.
 
 
\bibitem{Waxman1} E.\ Waxman and J.\ Bahcall, High Energy Neutrinos
from Astrophysical Sources: An Upper Bound, hep-ph/9807282.
 

\bibitem{Waxman2}  E.\ Waxman and J.\ Bahcall, \PRL 78 (1997) 2292.


 
\bibitem{Gaisser} T.D.\ Gaisser, F.\ Halzen and T.\ Stanev, \PRTS
258 (1995) 173.
 

\bibitem{Amelino1} G.\ Amelino-Camelia, J.\ Ellis, N.E.\ Mavromatos, 
D.V.\ Nanopoulos and S.\ Sarkar, \NAT
393 (1998) 763.
 


\bibitem{Biller} S.D.\ Biller et al., Limits to Quantum Gravity Effects
from
Observations of $TeV$ Flares in Active Galaxies,  gr-qc/9810044.




\bibitem{Ellis} J.\ Ellis, K.\ Farakos, N.E.\ Mavromatos, V.A.\ Mitsou and 
D.V.\ Nanopoulos, Astrophysical Probes of the Constancy of the
Velocity of Light,  astro-ph/9907340.
 
 
 
\bibitem{Colladay} D.\ Colladay and V.A.\ Kosteleck\'y, \PR
D55 (1997) 6760; D58 (1998) 116002. 



\bibitem{Coleman} S.\ Coleman and S.L.\ Glashow, \PL B405 (1997)
249; 
\PR D59 (1999) 116008. 



\bibitem{Phillips} P.R.\ Phillips, \PR 146 (1966) 966.



\bibitem{Nielsen} H.B.\ Nielsen and M.\ Ninomiya, \NP B141  
(1978) 153.


\bibitem{Weinberg} S.\ Weinberg, \PL
B138 (1984) 47.



\bibitem{Kostelecky1} V.A.\ Kosteleck\'y and S.\ Samuel, 
\PR D39 (1989) 683; \PRL 63 (1989) 224.



\bibitem{Bertolami1}  O.\ Bertolami, \CQG 14 (1997) 2748.



\bibitem{Bertolami2} O.\ Bertolami and D.F.\ Mota, \PL B455 (1999)
96. 



\bibitem{Kostelecky} V.A.\ Kosteleck\'y and R.\ Potting, \PR D51
(1995) 
3923; \PL B381 (1996) 389.



\bibitem{Dvali} G.\ Dvali and M.\ Shifman, Tilting the Brane, or Some
Cosmological Consequences of the Brane World, hep-th/9904021.



\bibitem{Kostelecky2} D.\ Colladay and V.A.\ Kosteleck\'y,
\PL B344 (1995) 259; \PR D52 (1995) 6224;

V.A.\ Kosteleck\'y and R.\ Van Kooten, \PR D54 (1996) 5585.



\bibitem{Bluhm1} R.\ Bluhm, V.A. Kosteleck\'y and N. Russell,
\PRL 79 (1997) 1432; \PR D57 (1998) 3932.



\bibitem{Bluhm2} R.\ Bluhm, V.A.\ Kosteleck\'y and N.\ Russell,
\PRL 82 (1999) 2254.



\bibitem{Bertolami3} O.\ Bertolami, D.\ Colladay, V.A.\ Kosteleck\'y
and R.\ Potting, \PL B395 (1997) 178.



\bibitem{Brillet} A.\ Brillet and J.L.\ Hall, \PRL 42 (1979) 549.



\bibitem{Hughes} V.W.\ Hughes, H.G.\ Robinson and V.\ Beltran-Lopez,
\PRL 4 (1960) 342.



\bibitem{Drever} R.W.P.\ Drever, \PMAG
6 (1961) 683.



\bibitem{Lamoreaux} S.K.\ Lamoreaux, J.P.\ Jacobs, B.R.\ Heckel, F.J.\  
Raab and E.N.\ Fortson, \PRL
57 (1986) 3125.


\bibitem{Kostelecky3} V.A.\ Kosteleck\'y and C.D.\ Lane, \PR D60
(1999) 
116010.


\bibitem{Bell} J.F.\ Bell, \AJ 462 (1996) 287.



\bibitem{BellD} J.F.\ Bell and T.\ Damour, \CQG
13 (1996) 3121.



\bibitem{Carroll} S.M.\ Carroll, G.B.\ Field and R.\ Jackiw, \PR D41
(1990) 1231;
S.M.\ Carroll and G.B.\ Field, \PRL 79 (1997) 2394.



\bibitem{Greisen} K.\ Greisen, \PRL
16 (1966) 748;

G.T.\ Zatsepin and V.A.\ Kuzmin, JETP Lett.\ 41 (1966) 78.



\bibitem{Halprin} A.\ Halprin and H.B.\ Kim, Mapping Lorentz Invariance
Violations into Equivalence Principle Violations, hep-ph/9905301. 



\bibitem{Mestres} L.\ Gonzales-Mestres, Deformed Lorentz Symmetry and 
Ultra-High Energy Cosmic Rays, hep-ph/9905430. 
 


\bibitem{Yoshida} N.\ Hayashida et al., (AGASA Collab.), \PRL 73 
(1994) 3491;
M.\ Takeda et al., (AGASA Collab.), \PRL 81 
(1998) 1163.



\bibitem{Bird} D.J.\ Bird et al., (Fly's Eye Collab.), \PRL 71
(1993)
3401; \AJ 424 (1994) 491; 441 (1995) 144.



\bibitem{Brooke}  M.A.\ Lawrence, R.J.O.\ Reid and A.A.\ Watson 
(Haverah Park Collab.), \JP G17 (1991) 733.



\bibitem{Efimov} N.N.\ Efimov et al., (Yakutsk Collab.), ICRR Symposium on
Astrophysical Aspects of the Most Energetic Cosmic Rays, eds.\ N.\ Nagano
and
F.\ Takahara, World Scientific, 1991.



\bibitem{Brucker} E.B.\ Brucker et al., \PR D34 (1986) 2183.


 
\bibitem{Glashow} S.L.\ Glashow, A.\ Halprin, P.I.\ Krastev, C.N.\ Leung
and 
J.\ Pantaleone, \PR D56 (1997) 2433.



\bibitem{Witten} E.\ Witten, \NP B471 (1996) 135.



\bibitem{Bertolami4}  O.\ Bertolami, \PL A154 (1991) 225.

\end{thebibliography}
\end{document}